**A hybrid helical structure of hard sphere packing from sequential deposition**

Ho-Kei Chan[1,2,#]

1. *Department of Physical and Theoretical Chemistry, School of Chemistry, University of Nottingham, University Park, Nottingham NG7 2RD, United Kingdom.*

2. *Foams and Complex Systems, School of Physics, Trinity College Dublin, College Green, Dublin 2, Ireland.*

\#    Email: epkeiyeah@yahoo.com.hk

**Abstract**

A hybrid helical structure of equal-sized hard spheres in cylindrical confinement was discovered as a 'by-product' of the recently developed sequential deposition approach [Physical Review E 84, 050302(R) (2011)] for constructing the densest possible packings of such systems. Unlike the conventional triple-helix structure where its three strands of spheres are packed densely to form triads of close-packed, mutually touching spheres, in this novel helical phase only two of its three strands of spheres are packed in this densest arrangement and the overall structure resembles a hybrid of the single and the double helix. This article explains how this previously unknown structure can be constructed via the abovementioned sequential deposition of spheres, which involves manipulating the positions of a few spheres to create a template for the deposition process. The findings show that it is possible to discover new structures through varying only the configuration of few spheres within the template, where this approach relies on a sensitive dependence of the deposition-generated structures on the template.

**Introduction**

Dense packings of equal-sized hard spheres in cylindrical confinement could serve as a model for a variety of physical systems with spherical entities [1-6], such as colloidal crystal wires [1] and fullerenes contained inside nanotubes [2]. For such systems a variety of densely packed structures have been observed as a function of the diameter ratio $D$ = cylinder diameter / sphere diameter (i.e. cylinder diameter in units of the sphere diameter). It was found that these structures exhibit structural similarity [1,2] with the computationally predicted densest possible structures for equal-sized hard spheres in cylindrical confinement [7-14]. In previous work, simulated annealing has been employed [7-10] to predict such densest possible structures of hard spheres up to at least $D = 2.873$ [10], and a corresponding method of sequential deposition [11] has been developed for constructing such densest possible structures up to $D = 2.7013$ (where below this value of $D$ the densest possible structures consist only of spheres that are in touch with the cylinder). The general problem of packing equal-sized spheres into a cylinder has also been studied in other contexts [15-18], where in some cases the idea of sequential deposition [19] was employed [15]. For the problem addressed here, the densest possible configurations as predicted from simulated annealing [7-10] include the peapod-like "bamboo" structure [$D = 1$, Fig. 1(a)], the normal zigzag (also referred to as the planar zigzag) [$1 < D \leq 1.866$, Fig. 1(b)], the single helix [$1.866 < D \leq 1.99$, Fig. 1(c)], the double helix [$1.99 < D < 2$, Fig. 1(d)], and the achiral doublets [$D = 2$, Fig. 1(e)], as well as a range of multiple-helix structures (Fig. 2) [7,9,10] that have also been referred to as line-slip [9,10] or line-defect [11] structures. As shown from Figs. 2(a) to 2(d), a multiple-helix structure is characterized by a repeating multiple strand of spheres (single strand for single helix, double strand for double helix, etc.) on its surface pattern. For $D \geq 2$, the aforementioned method of sequential deposition involves a continuous variation of the underlying template [11] (i.e. the configuration of spheres at the base of a vertically aligned cylinder) for finding the densest possible structure. In such a variational search, some additional ordered periodic structures were found. Most of them have been expected from theory [9,10], but among them was a previously unknown structure that exists in the form of a hybrid of the single and the double helix, between $D \approx 2.22$ and $D \approx 2.29$. The surface pattern of this unique helical phase consists of a repeating triple strand of spheres. But unlike the triple-helix structure [Fig. 2(c)] where its three strands of spheres are all packed densely to form triads of close-packed, mutually touching spheres [as indicated by a triangle in Figs. 2(c) and 2(d)], in this novel helical phase only two of its three strands of spheres are packed in this densest arrangement and the pair of strands is packed at a lower density with the remaining strand. As a result, the overall structure as well as its surface pattern resemble a hybrid of the single- and the double-helix structure. This article aims to give a full account of this unique phase, including information on how it can be constructed via a sequential deposition of spheres.

**Method of sequential deposition**

In this section, the abovementioned method of sequential deposition is reviewed with some pictorial explanations that are complementary to the original description in Ref. [11]. Consider cases of $1 \leq D \leq 2.7013$ where the densest possible structures consist only of spheres that are in touch with the cylindrical container. The main idea is to deposit the spheres one by one onto their lowest possible (geometric) positions inside a vertically aligned cylindrical tube. Take the normal zigzag at $D = 1.3$ as an example: Starting with a first sphere on the flat base of the cylinder (Fig. 3a), the lowest possible position of each subsequent sphere corresponds to an angular separation of $180°$ from the position of the previous sphere. By placing each subsequent sphere in such a lowest possible position, the normal zigzag structure can be obtained, as shown in Fig. 3.

The problem of degeneracy, i.e. the existence of multiple lowest possible positions, is resolved by defining a direction in the angular scan from the position of the previous sphere for lowest possible positions. This direction can be counterclockwise or clockwise, corresponding respectively to an increase or decrease in the value of the angular position $\phi$. Whenever multiple lowest possible positions exist, the first lowest possible position encountered in the angular scan is chosen.

The flat base of the cylinder, as depicted in the example in Fig. 3, leads to agreement with the densest possible structures predicted from the method of simulated annealing *only* for the regime of $1 \leq D \leq 2$. The reason is that, for $D \geq 2$, the lowest possible position of the second sphere (and also some other spheres for larger $D$ values) is on the flat base of the cylinder (Fig. 4). This implies that subsequent spheres, which are to be placed at their lowest possible positions (often above some other spheres), might not be placed at "proper" positions (vertical or angular) that correspond to the densest possible packings. The remedy for this is the creation of a multi-sphere template on the flat base of the cylinder. This template mimics the end of a sphere column, such that subsequent spheres in the deposition process could be "shifted" into "proper" vertical or angular positions that yield the densest possible packings. Note that in the finished template, the positions of the few spheres within the template are fixed (by whatever means), and no subsequent sphere can be dropped onto the cylinder's base. The template is created in the spirit to preserve a continuity in relative positions between successive spheres: for any chosen values of the vertical separation $(\Delta z)_{2,1}$ (in units of the sphere diameter) and angular separation $(\Delta\phi)_{2,1}$ between the first and second spheres, the $N^{th}$ sphere of the template is to be placed at the vertical position $(N-1)(\Delta z)_{2,1}$ and angular position $(N-1)(\Delta\phi)_{2,1}$, both relative to the position of the first sphere (Fig. 5(a)). This rule is followed regardless of whether the first two spheres are in touch with each other (Fig. 5(b)) or not (Fig. 5(a)). However, if a sphere can be placed at a lower position such that it would come into contact with a sphere placed earlier, (Fig. 5(c)), it will be placed at that lower position as an exception to the continuity rule described above.

The parameters $(\Delta z)_{2,1}$ and $(\Delta\phi)_{2,1}$ are not always chosen separately. For any given angular separation $(\Delta\phi)_{2,1}$, the vertical separation $(\Delta z)_{2,1}$ is initially chosen to be the lowest possible vertical separation between the two spheres. For $1 \leq D < 2$, the second sphere must be in contact with the first (Fig. 4), and therefore the vertical separation $(\Delta z)_{2,1}$ can simply be determined from Eq. (1) of Ref. [11],

$$(\Delta z)^2 + \frac{(D-1)^2}{2}[1-\cos\Delta\phi] = 1 \tag{1}$$

which describes two cylinder-touching spheres in mutual contact. For $\Delta z = (\Delta z)_{2,1}$ and $\Delta \phi = (\Delta \phi)_{2,1}$, it follows that

$$(\Delta z)_{2,1} = \sqrt{1 - \frac{(D-1)^2}{2}[1 - \cos(\Delta \phi)_{2,1}]} \qquad (2)$$

for *any* value of $(\Delta \phi)_{2,1}$.

For $D \geq 2$, the rule is slightly different: If the magnitude of $(\Delta \phi)_{2,1}$ is smaller than or equal to a critical angular separation $(\Delta \phi)_{2,1}*$ such that the second sphere can be lowered vertically and brought into contact with the first sphere, the second sphere will be placed at that lowest possible position. Else if it is $(\Delta \phi)_{2,1} > (\Delta \phi)_{2,1}*$ such that $(\Delta z)_{2,1} = 0$, i.e. if the second sphere falls onto the cylinder's flat base without coming into contact with the first sphere, the value of $(\Delta z)_{2,1}$ has to be assigned separately. Note that the value of $(\Delta \phi)_{2,1}*$ corresponds to a situation of $(\Delta z)_{2,1} = 0$ where the second sphere is in contact with the first one, which implies

$$(\Delta \phi)_{2,1}* = \cos^{-1}\left\{1 - \frac{2}{(D-1)^2}\right\} \qquad (3)$$

according to Eq. (2).

For $1 \leq D < 2$, the densest possible structure is chosen only from a continuous variation of the angular separation $(\Delta\phi)_{2,1}$, as the vertical separation $(\Delta z)_{2,1}$ would always be determined from Eq. (1). And for $D \geq 2$, apart from a continuous variation of $(\Delta\phi)_{2,1}$, there would also be a continuous variation of the vertical separation $(\Delta z)_{2,1}$ if $(\Delta\phi)_{2,1} > (\Delta\phi)_{2,1}*$. Whenever a sphere is to be deposited towards a position lower than that of any previous sphere, the final structure of spheres is discarded and the search for densest possible packings continues with other values of $(\Delta\phi)_{2,1}$ and $(\Delta z)_{2,1}$. This computational "shortcut", which was not explicitly mentioned in Ref. [11], was implemented to achieve a faster search algorithm for the densest possible structures being studied.

**Results: the single-double hybrid helix**

In searching for the densest possible configuration at any diameter ratio $D$, the volume fraction, $V_F \equiv$ volume of spheres / volume of cylinder, is investigated numerically as a function of the angular separation $(\Delta\phi)_{2,1}$, and also of the vertical separation $(\Delta z)_{2,1}$ if $(\Delta \phi)_{2,1} > (\Delta \phi)_{2,1}{}^*$. For cases of $(\Delta \phi)_{2,1} \leq (\Delta \phi)_{2,1}{}^*$ at $D > 2$, the plot of $V_F$ against $(\Delta \phi)_{2,1}$ consists typically of two peaks, where each peak corresponds to a particular locally or globally densest multiple-helix structure characterized by a repeating multiple strand on its surface pattern with one of these peaks corresponding to the densest possible packing as predicted from simulated annealing. An example of this is the triple helix and quadruple helix at $D = 2.35$ (Fig. 6). Yet, the scenario is slightly different for values of $D$ between 2.22 and 2.29: For the regime $(\Delta \phi)_{2,1} \leq (\Delta \phi)_{2,1}{}^*$, there are still two peaks of locally densest structures in the plot of $V_F$ against $(\Delta \phi)_{2,1}$ (Fig. 7). But from $D \approx 2.22$ to $D \approx 2.26$ neither of these peaks corresponds to the densest possible structure, which turns out to be a double helix (structure 14 in Fig. 8 of Ref. [10]) that has been constructed from sequential deposition at $(\Delta \phi)_{2,1} > (\Delta \phi)_{2,1}{}^*$. And more significantly, for the whole regime from $D \approx 2.22$ to $D \approx 2.29$, the peak at a lower value of $(\Delta \phi)_{2,1}$ does not correspond to one of our familiar multiple-helix structures, but to a previously unknown helical structure which presents itself like a hybrid of the single- and the double-helix structure [Fig. 8(a1) to 8(a7)] - In the repeating triple strand of this new structure, two of its three strands are packed densely to form triads of close-packed, mutually touching spheres but the pair of strands is packed at a lower density with the remaining strand. The peak at a higher value of $(\Delta \phi)_{2,1}$ corresponds to the conventional triple-helix structure

(see Fig. 8(b1) to 8(b7)), where all of its three strands are packed in a densest configuration (i.e. hexagonal-like on its surface patterns), as different from the case of the hybrid helix. As shown in Fig. 8, the qualitative difference between these two helical structures is clear from $D \approx 2.22$ to $D \approx 2.25$, but fades out for an increase of $D$ from around 2.26 to 2.29. At $D \approx 2.29$, the two helical structures 'merge' into a uniform, defect-free structure (Fig. 8(c)), which is known as the "431 symmetric" structure in a recently developed phyllotactic description [9,10] for such packings. Such a decrease in structural difference as $D$ increases towards 2.29 is quantitatively depicted in Figs. 7 and 9, from which a decreasing difference in $V_F$ and $(\Delta\phi)_{2,1}$ between the two structures can be seen.

Using the case of $D = 2.22$ as an example, Fig. 10 illustrates how the hybrid-helix structure can be constructed via the aforementioned method of sequential deposition. It is worth noting that the third and the fourth sphere in the hybrid helix are located on the same vertical position. For both the hybrid- and the triple-helix structure, the template is composed of the first three spheres only, since these three spheres can already prevent any sphere from being dropped onto the cylinder's base. The helical structure of spheres is then generated via the method of sequential deposition described above. Since $(\Delta\phi)_{2,1} < (\Delta\phi)_{2,1}*$ and $(\Delta\phi)_{3,1} = 2 \times (\Delta\phi)_{2,1}$

[where $(\Delta\phi)_{3,1}$ is the angular separation between the first and third sphere], only the angular separation $(\Delta\phi)_{2,1}$ between the first two spheres needs to be specified for creating the right template. Table 1 shows the numerical values of $(\Delta\phi)_{2,1}$ for both helical structures at various values of $D$ between 2.22 and 2.29.

**Conclusions**

The findings reported here show that it is possible to discover new structures for such quasi-1D systems through varying the configuration of spheres within the underlying template, where this approach relies on a sensitive dependence of the deposition-generated structures on the template. The template dependence of structures formed from a deposition process is encountered not only in this sphere-packing problem but also on other occasions like the experimental realization of the Weaire-Phelan foam structure [20]. While such sensitivity to the underlying template might be a drawback for controlling structures formed in experimental processes, it opens up possibilities of discovering new structures like the one presented in this article. It would be interesting to extend the search for new structures to other values of $D$ and also to templates that are different from the ones developed in the abovementioned sequential deposition approach.

**Acknowledgements**

This work was supported by the Irish Research Council through an EMPOWER Post-doctoral Fellowship (Dec 2010 - Dec 2012). Prof. Tony Stace and Dr. Alexander Markevich, both from the School of Chemistry of the University of Nottingham, have kindly helped to proofread the original manuscript. Indispensible support from Chloe Po-Yee Wong at a time of uncertain future is gratefully acknowledged.

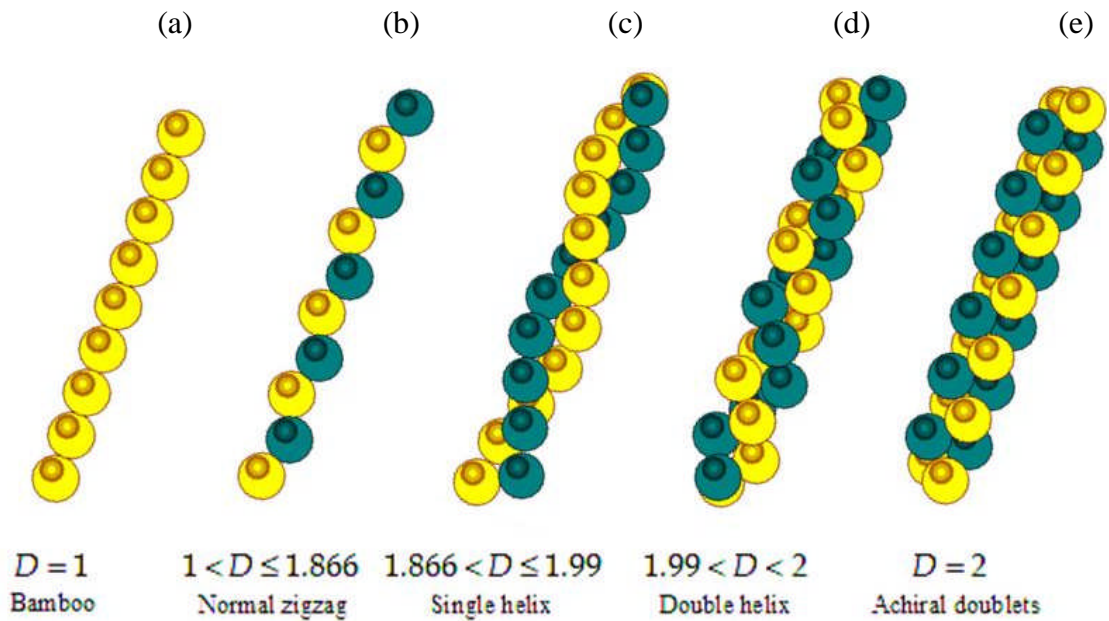

*Figure 1    The first five computationally predicted densest possible structures [8-10] for equal-sized hard spheres in cylindrical confinement, at various regimes of the diameter ratio D = cylinder diameter / sphere diameter. Two different colours are used to illustrate the nature of each structure: Both the single- and the double-helix consist of an intertwisting pair of strands of spheres, where the structure of achiral doublets has every two spheres located on one vertical position.*

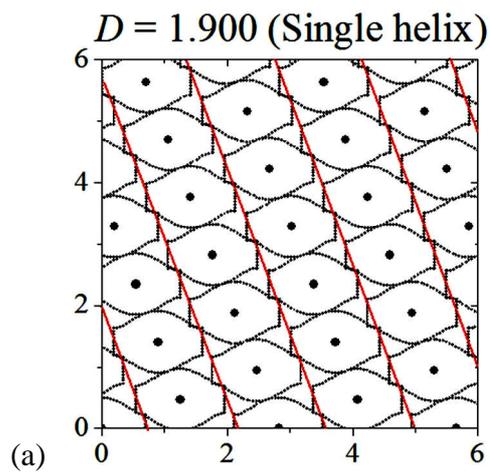
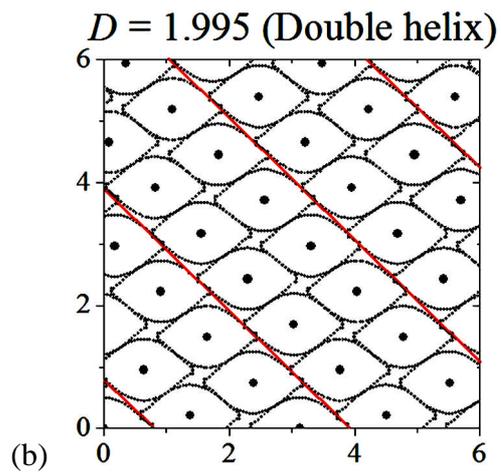
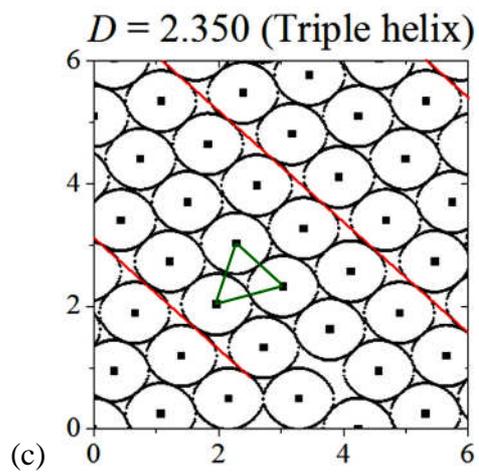
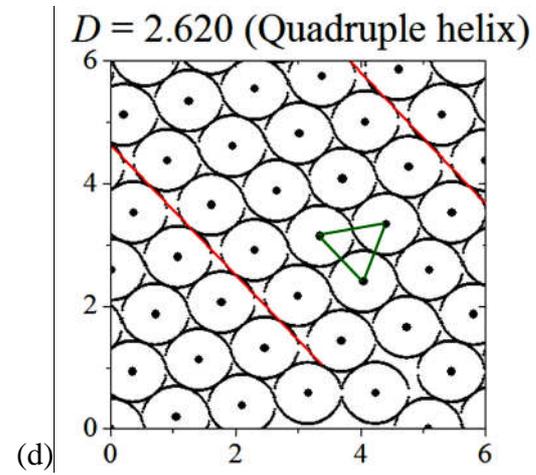

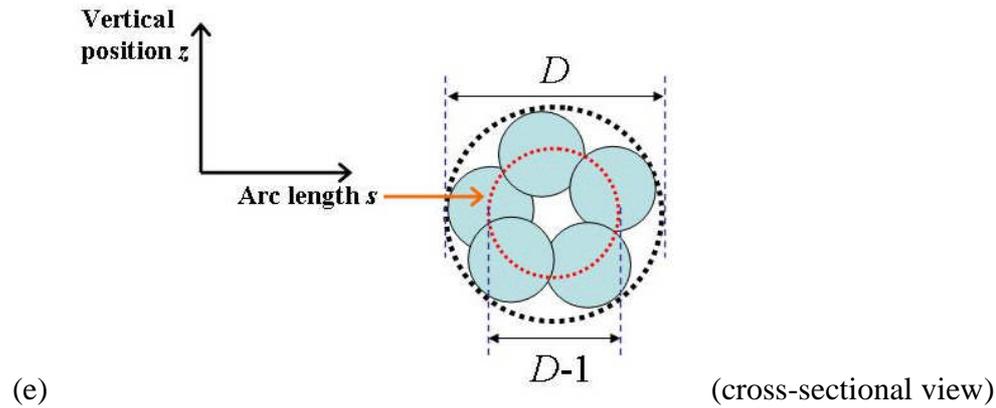

(e)  (cross-sectional view)

*Figure 2    Surface patterns [11] of the single-, double-, triple- and quadruple-helix densest possible structures at (a) D = 1.900, (b) D = 1.995, (c) D = 2.35 and (d) D = 2.620, respectively. The solid lines are used as a guide to indicate the repeating single, double, triple or quadruple strand of spheres in the surface patterns, and in (c) and (d) a triangle is used to indicate a triad of close-packed, mutually touching spheres. (e) Explanation of the surface patterns shown from (a) to (d): they are spatial distributions of sphere centres on the z-s plane, where z is the position coordinate along the cylinder's axis and s is the coordinate associated with arc lengths of the "inner" cylindrical surface (see figure) that embeds all of the sphere centres. Note that the boundary of each sphere on such surface plots, which represents a particular kind of projection onto the curved surface of the cylinder, approaches a circular shape only if the diameter ratio D is well above 2 (i.e. only if the cylindrical surface is not too curved). Ref. [11] contains an explanation of how those boundaries of spheres (e.g. oval- or "peanut"-shaped) on the surface plots are obtained.*

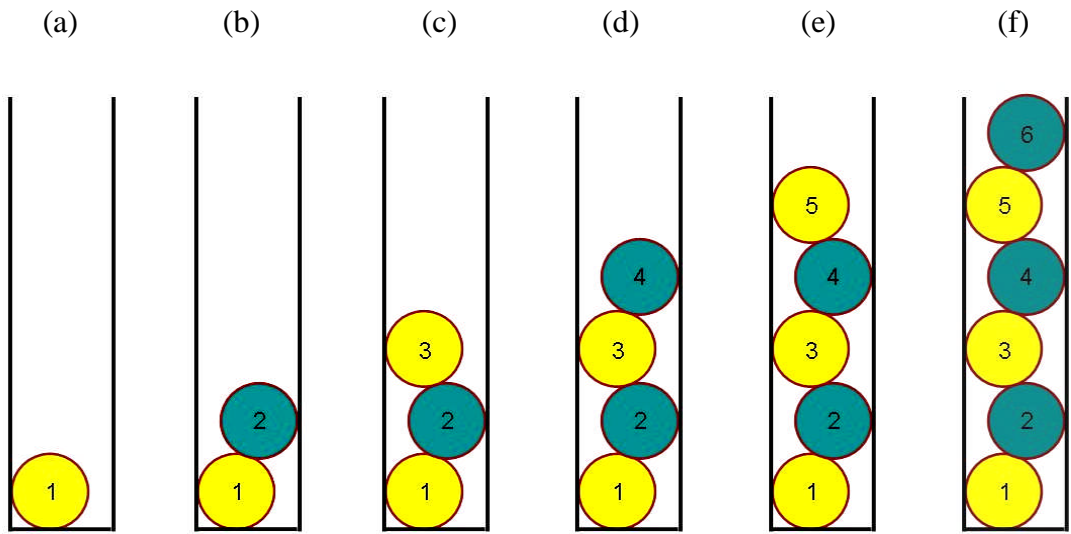

*Figure 3     Construction of the normal zigzag structure at D = 1.3 via sequential deposition. Initially, a sphere is placed on the flat base of the cylinder. At each subsequent deposition step, a sphere is placed at its lowest possible position, which corresponds to an angular separation of 180° from the previous sphere. It follows that all the spheres are residing on the same plane, hence this structure is also referred to as "planar zigzag".*

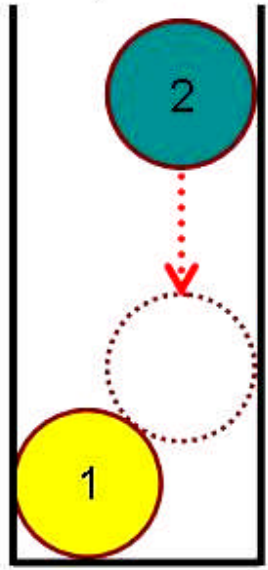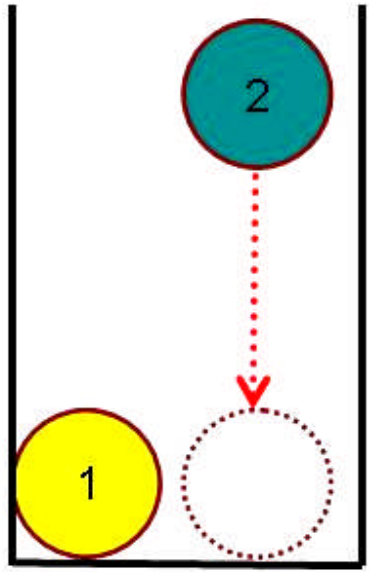

*Figure 4      At D ≥ 2, the second sphere can be dropped onto the flat base of the cylinder, which is not the case for D < 2.*

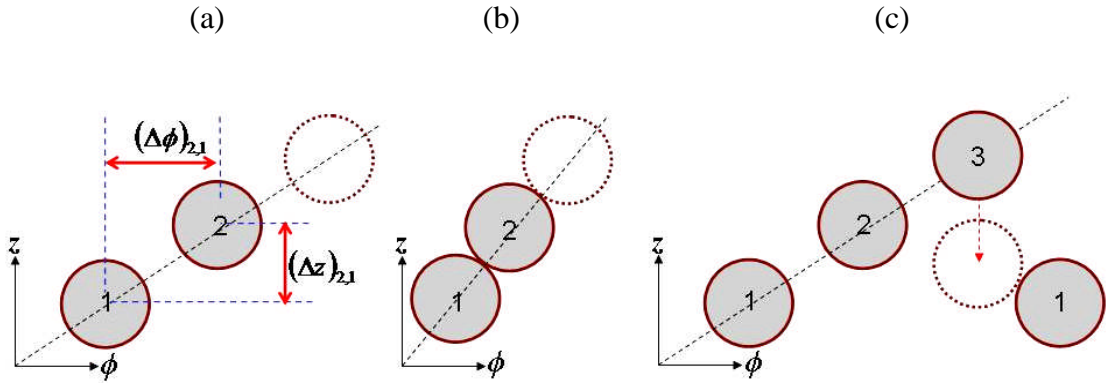

*Figure 5     The method of template creation generally follows a continuity in the relative positions between successive spheres in the template, as illustrated by the dotted lines. (a) The angular separation $(\Delta\phi)_{2,1}$ between the first two spheres is chosen first, followed by a separate choice of the vertical separation $(\Delta z)_{2,1}$ if at that chosen angular separation the second sphere cannot be brought into contact with the first one. (b) $(\Delta z)_{2,1}$ would be the minimum possible vertical separation between the first and second spheres if at the chosen angular separation the second sphere can be vertically brought into contact with the first sphere, i.e. if $(\Delta\phi)_{2,1} \leq (\Delta\phi)_{2,1}*$. (c) As an exception to the continuity rule, if at the chosen angular separation a sphere can be placed at a lower position such that it would come into contact with an underlying sphere, it will be placed at that lower position.*

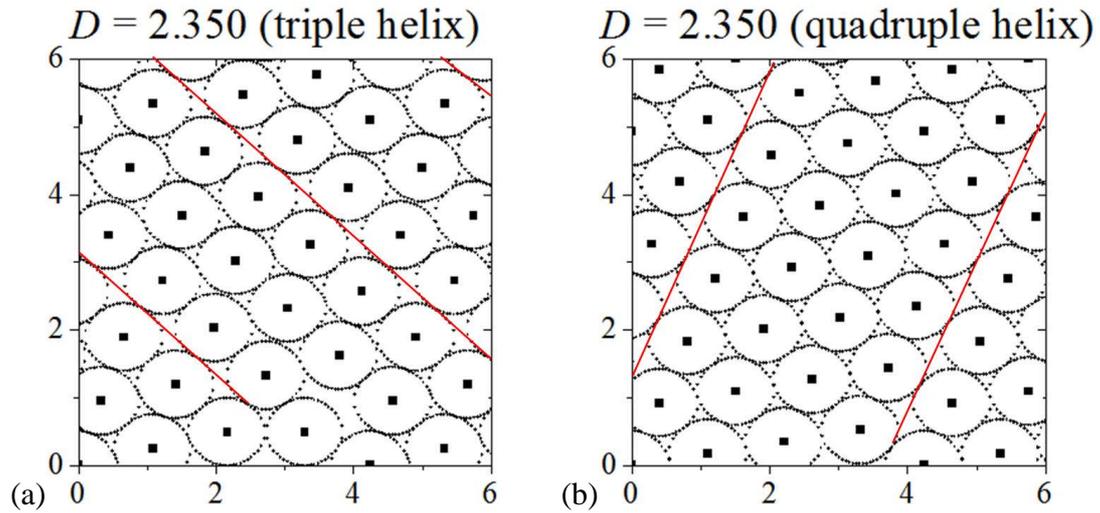

*Figure 6    Surface patterns of two helical structures at D = 2.35: (a) The densest possible triple-helix structure at $(\Delta\phi)_{2,1}$ = 91.8576° [same as Fig. 2(c)], and (b) the less dense, quadruple-helix structure at $(\Delta\phi)_{2,1}$ = 93.7152°. The solid lines are used as a guide to indicate the repeating triple or quadruple strand of spheres in the surface patterns. Note the sensitive dependence of the final structure of spheres on the value of $(\Delta\phi)_{2,1}$.*

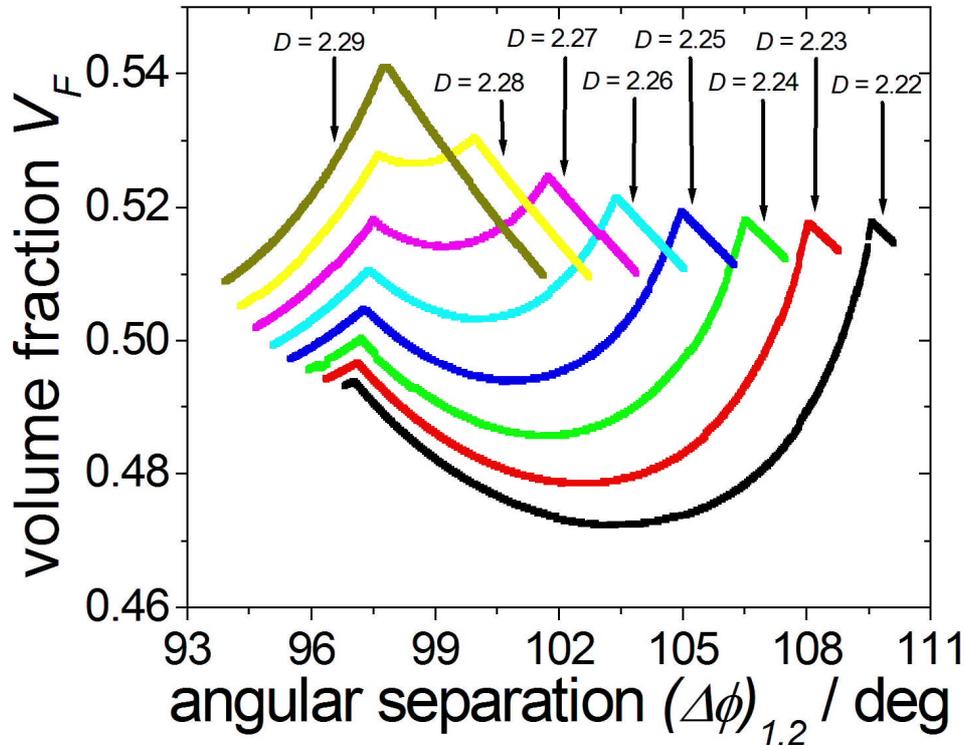

*Figure 7* Plot of the volume fraction $V_F$ against the angular separation $(\Delta\phi)_{2,1}$ for values of D between 2.22 and 2.29. Each curve is characterized by two local peaks: The peak at a lower value of $(\Delta\phi)_{2,1}$ corresponds to the hybrid-helix structure, where the peak at a higher value $(\Delta\phi)_{2,1}$ corresponds to the triple-helix structure. As D increases towards 2.29, the two peaks approach each other both along the horizontal and the vertical axis, corresponding to a decreasing difference between the two structures. At $D \approx 2.29$, only one peak is left, meaning that the two helical structures have "merged into" a single defect-free structure.

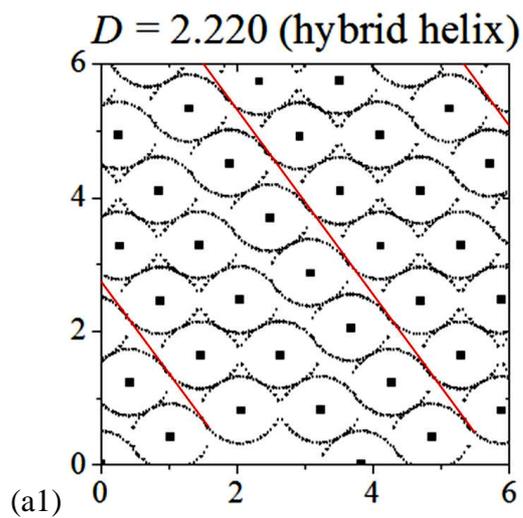
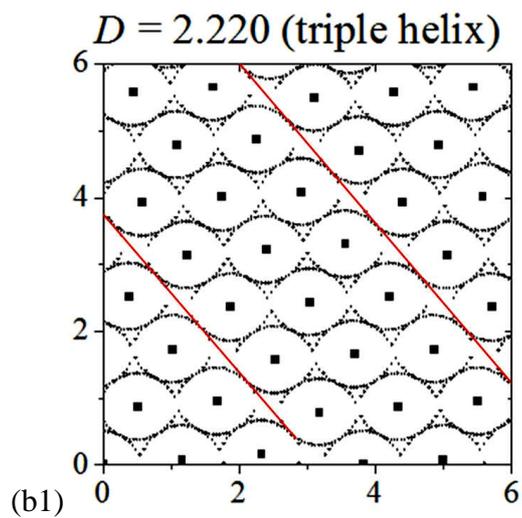
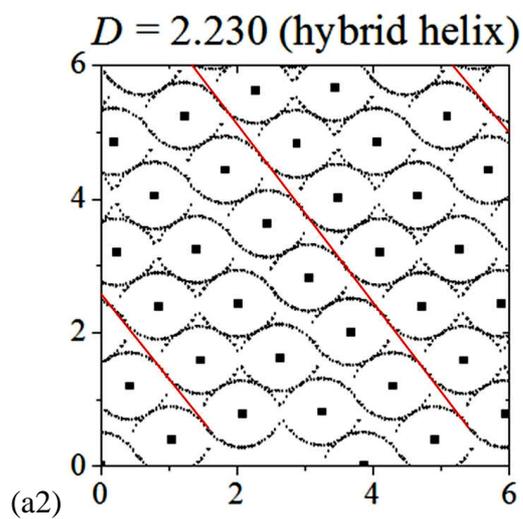
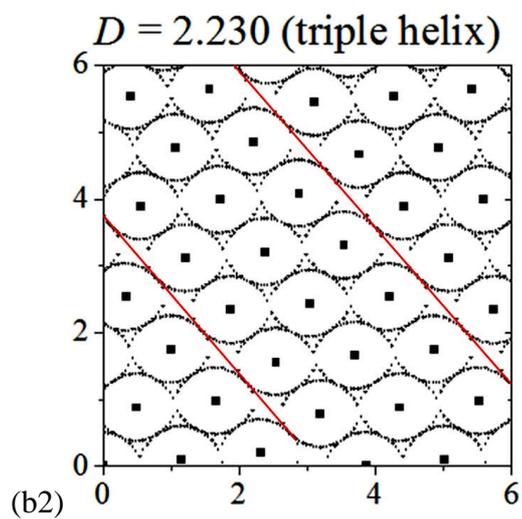
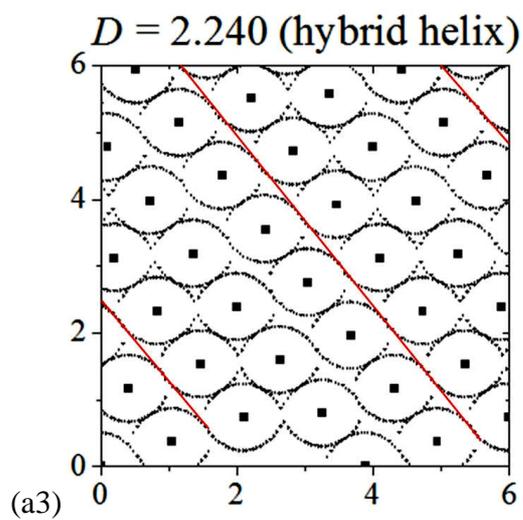
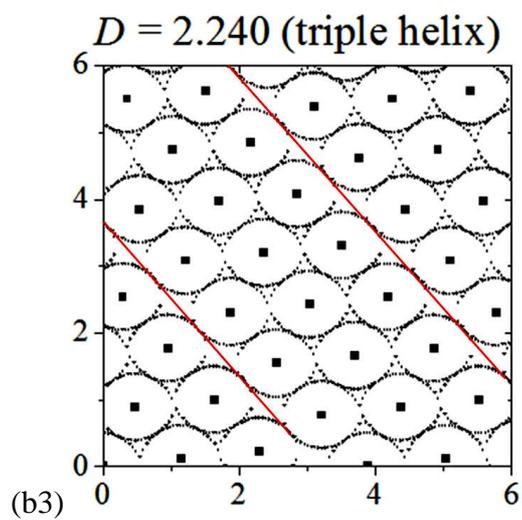

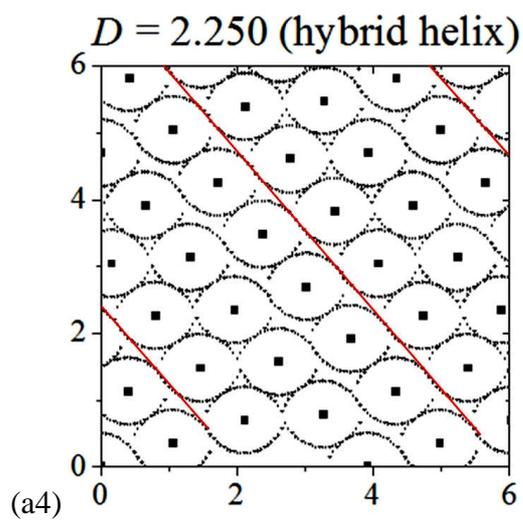
(a4)

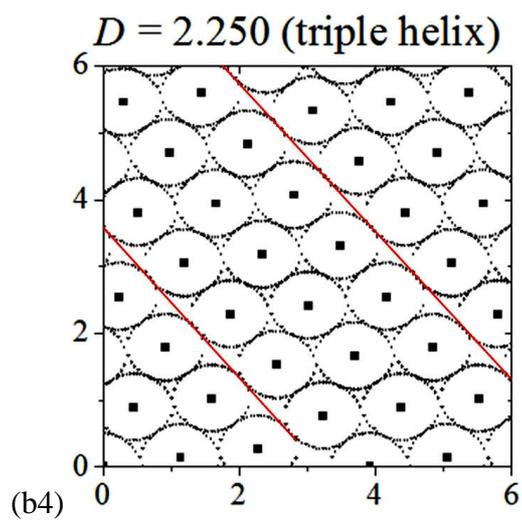
(b4)

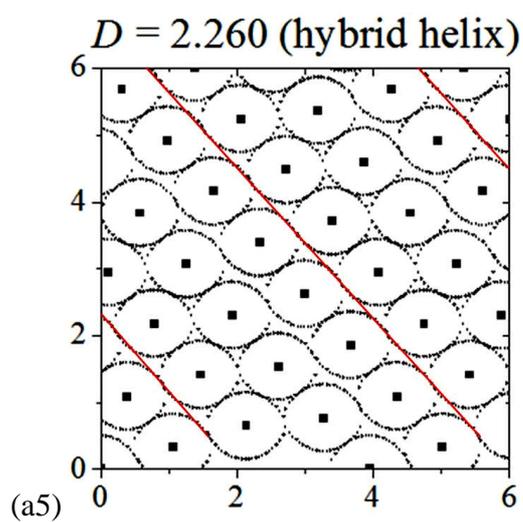
(a5)

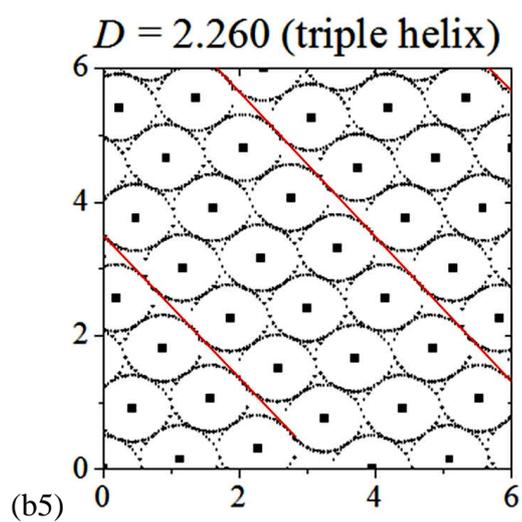
(b5)

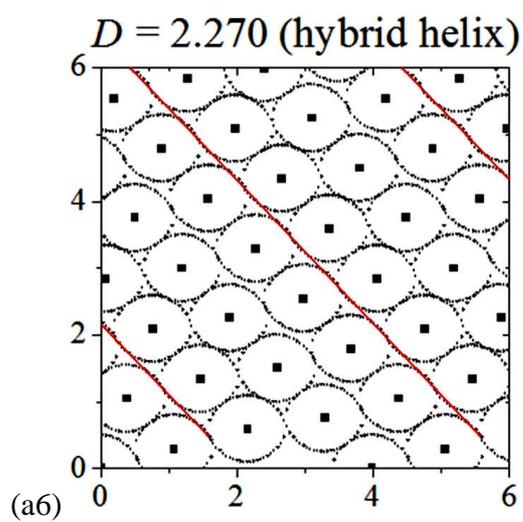
(a6)

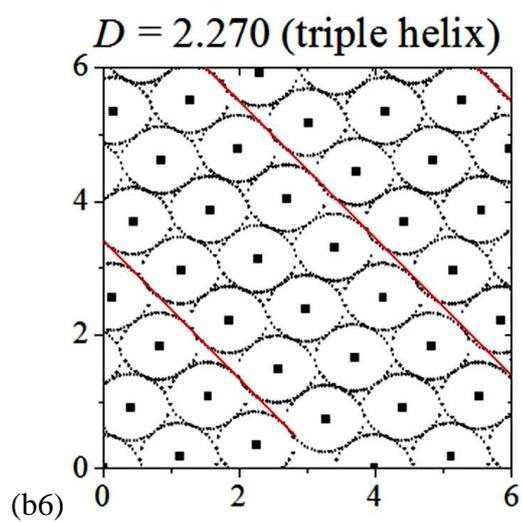
(b6)

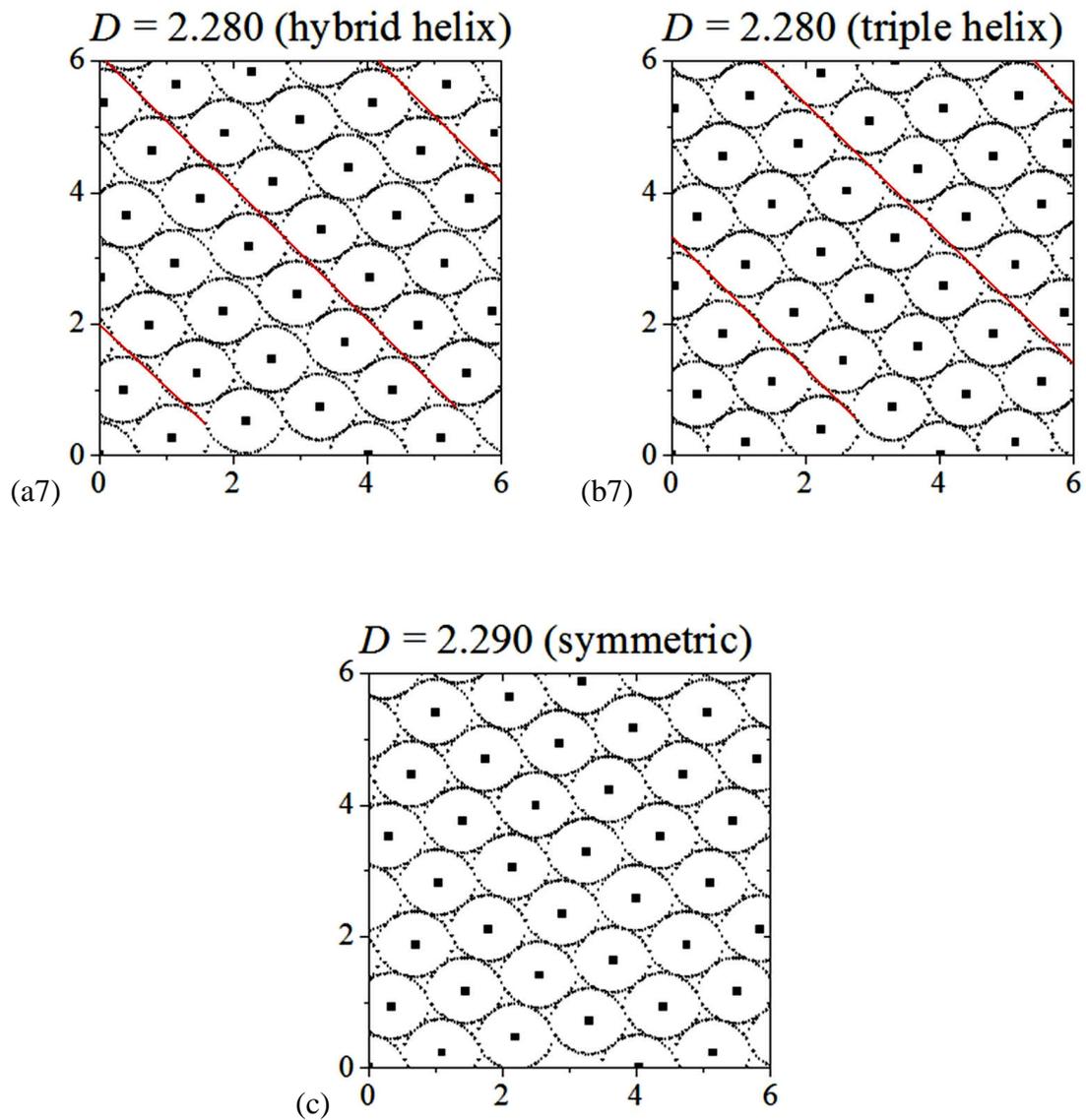

*Figure 8 Surface patterns of (a) the hybrid helix and (b) triple helix, from D ≈ 2.22 to D ≈ 2.28. The qualitative difference between the two structures is clear from D ≈ 2.22 to D ≈ 2.25, but starts to fade out as D increases from 2.26 towards 2.29. (c) At D ≈ 2.29, the system is only left with a single defect-free structure. The solid lines are used as a guide to indicate the repeating triple strand of spheres in the surface patterns.*

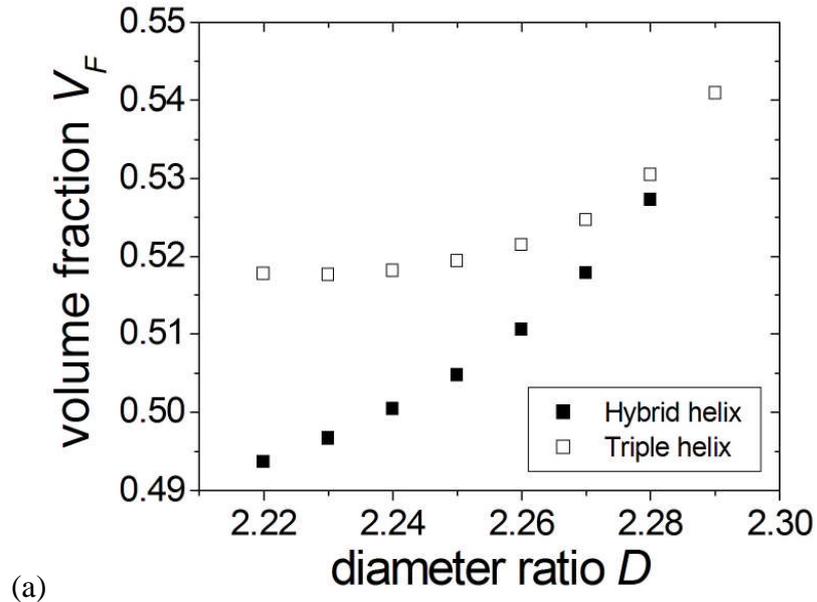

(a)

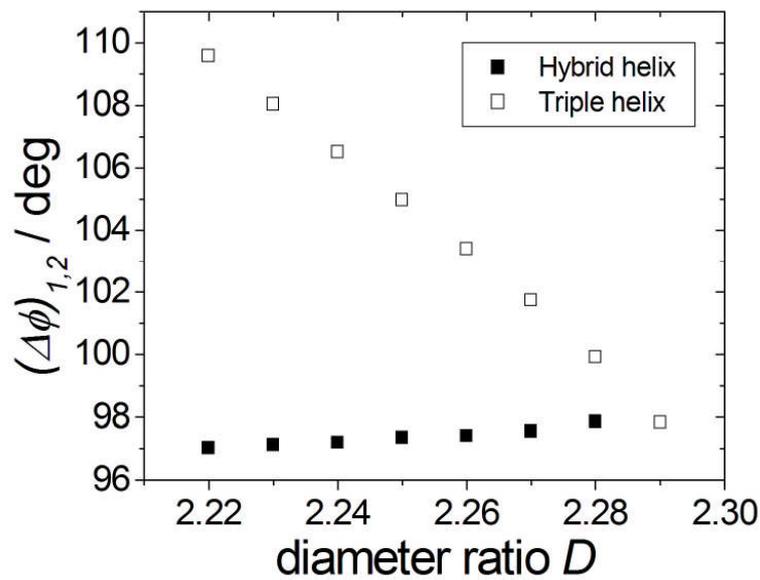

(b)

*Figure 9    Plot of the (a) volume fraction $V_F$ as well as (b) the angular separation $(\Delta\phi)_{2,1}$ against the diameter ratio D, for both helical structures. As D increases towards approximately 2.29, these two quantities approach each other, corresponding to a decreasing difference between the two helical structures.*

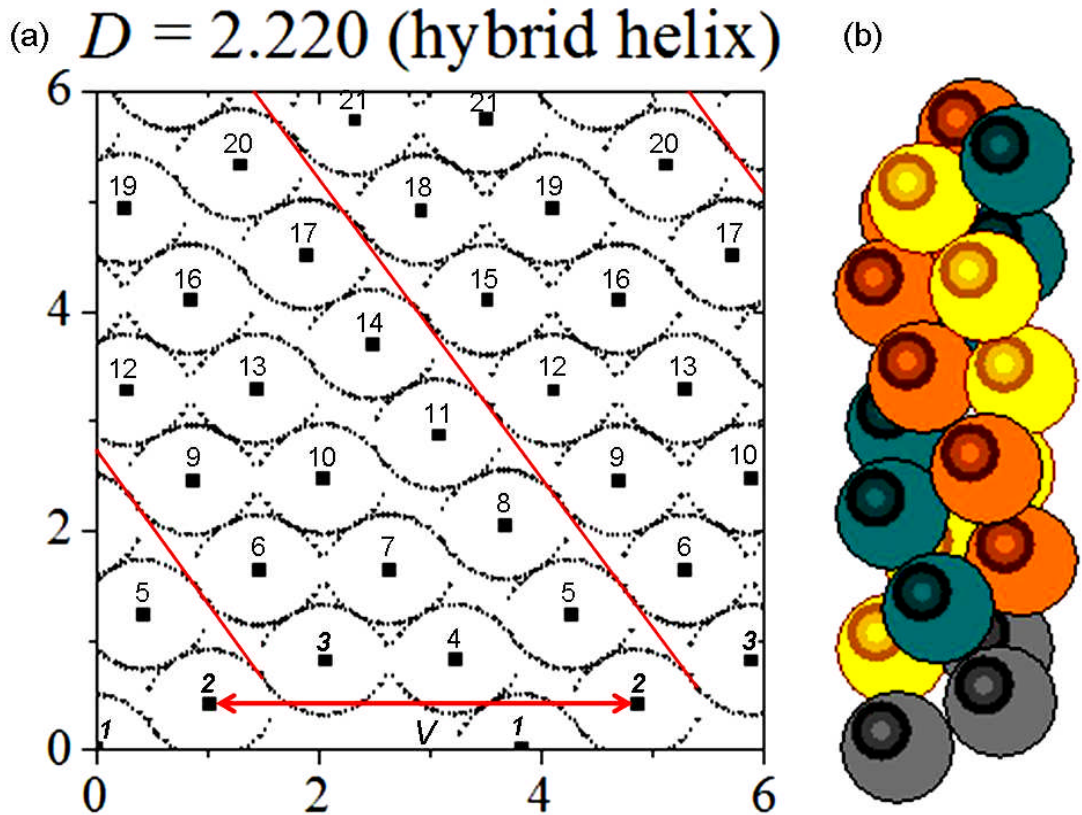

*Figure 10 (a) Illustration of the sequential deposition procedure for the hybrid-helix structure at D = 2.22. The numbers on the surface pattern denote the sequence of sphere deposition, and the numbers in italics denote the spheres that form the template. The periodicity V along the horizontal axis corresponds to the circumference of the "inner" cylindrical surface (Fig. 2(c)). The solid lines are used as a guide to indicate the repeating triple strand of spheres in the surface patterns. (b) A 3-D image of this structure. The first three spheres form the template. Note that for this particular case of D = 2.22 the third and the fourth sphere are located on the same vertical position.*

| Diameter ratio $D$ | $(\Delta\phi)_{2,1}$ (Hybrid helix) | $(\Delta\phi)_{2,1}$ (Triple helix) |
|---|---|---|
| 2.22 | 97.0272° | 109.577° |
| 2.23 | 97.128° | 108.043° |
| 2.24 | 97.2° | 106.51° |
| 2.25 | 97.3576° | 104.976° |
| 2.26 | 97.416° | 103.406° |
| 2.27 | 97.5528° | 101.736° |
| 2.28 | 97.8696° | 99.9216° |

*Table 1 Numerical values of $(\Delta\phi)_{2,1}$ for the plot in Fig. 9(b).*